\documentclass[twocolumn,prd]{revtex4}

\usepackage{amsmath}
\usepackage{graphicx}
\usepackage{epstopdf}
\usepackage{colordvi}
\usepackage{url}
\newcommand{\beq}{\begin{equation}}
\newcommand{\eeq}{\end{equation}}

\begin{document}
%
\title{Noise Estimate of Pendular Fabry-Perot through Reflectivity Change}
\author{Paolo Addesso}
\affiliation{Dept. of Electrical Eng. and Information Eng.
 University of Salerno
Via Ponte Don Melillo 1, 84084 Fisciano, IT}
\email[Email:]{paddesso@unisa.it}
\author{Vincenzo Pierro}
\homepage[url:]{http://scholar.google.it/citations?user=80kLutcAAAAJ}
\affiliation{ Dept. of Engineering
University of Sannio
C.so Garibaldi 107, 82100 Benevento, IT}
\email[Email:]{pierro@unisannio.it}
\author{Giovanni Filatrella}
\affiliation{
Dept. of Sciences and Technology 
 University of Sannio Via Port'Arsa 11, 82100 Benevento, IT}
\email[Email:]{filatrella@unisannio.it }

\begin{abstract}
A key issue in developing pendular Fabry-Perot interferometers as very accurate displacement measurement devices, is the noise level. 
The Fabry-Perot pendulums are the most promising device to detect gravitational waves, and therefore the background and the internal  noise should be accurately measured and reduced.
In fact terminal masses generates additional internal noise mainly due to thermal fluctuations and vibrations.
We propose to exploit the reflectivity change, that occurs in some special points, to monitor the pendulums  free oscillations and possibly estimate the noise level. 
We find that in spite of long transients, it is an effective method for noise estimate.
We also prove that to only retain the sequence of escapes, rather than the whole time dependent dynamics, entails the main characteristics of the phenomenon.
Escape times could also be relevant for  future gravitational wave detector developments.
\end{abstract}

\maketitle

\section{Introduction}

Noise at thermal equilibrium is given by the temperature, i.e. it is connected to the spontaneous flow of energy. 
In a macroscopic device, where the spontaneous flow of energy often cannot be even defined \cite{Risken89}, random perturbations of the dynamical variables can be characterized by some equivalent noise level, by measuring, say, the standard deviation of a representative variable. 
For instance the noise level in an electronic device can be measured detecting  current (or voltage) fluctuations and measuring the background by the power spectral density. 

The problem is very complicated for low noise devices, where the measurement itself changes the
properties of the devices \cite{Bodiya12} \cite{Villar10} (see the case of quantum noise characterization \cite{Chan11}\cite{Ludwig08}), and could introduce a substantial amount
of extra disturbances. 

A preeminent example is the measurement of noise levels in  pendular Fabry-Perot (henceforth FP) interferometers,  that is the most promising (and intriguing \cite{Aguirregabiria87}\cite{Pierro94}) candidate detection system for gravitational waves \cite{Deruelle84}.
This devices consist (see Ref. \cite{Rakhmanov98} for details) of multiple suspended pendulums to reduce the external noise , and terminal masses far apart to make the optical distance of several kilometers (in order to enhance sensitivity and enlarges the useful frequency band \cite{InterfeWeb}). 
Still, it is required that the macroscopic object are subject to background noise as low as to achieve (in the next future) quantum fluctuations
of the macroscopic mass.
A possibility to measure the noise level is to indirectly retrieve the fluctuations. An indirect effect of fluctuations is given by the Kramers escape \cite{Kramers40}. 
It is well known that a particle (or, in general, a degree of freedom) subject to white noise of intensity (correlator, in statistical terms) 
$D$ and trapped in a potential of height $\Delta U$ and subject to a friction $\gamma$ overcomes the barrier at a rate:
\beq
r = r_0 \exp\left(- \frac{\gamma \Delta U}{D}\right)
\label{kramers}
\eeq
It is therefore tempting to only measure the rate $r$ detecting the passages through special points related to the barrier $\Delta U$. 
From the point of view of accuracy, the exponential dependence in Eq.(\ref{kramers}) is promising, in that small variations of the noise intensity $D$ induce large changes in the escape rate $r$. 
We show that this is in fact the case: despite we reduce the sampled dynamics to relatively few points, the noise intensity estimate is rather effective,
 and the reliability can be analytically estimate as a function of the system parameters.

\begin{figure}[!t]
\centering
\includegraphics[width=2.5in]{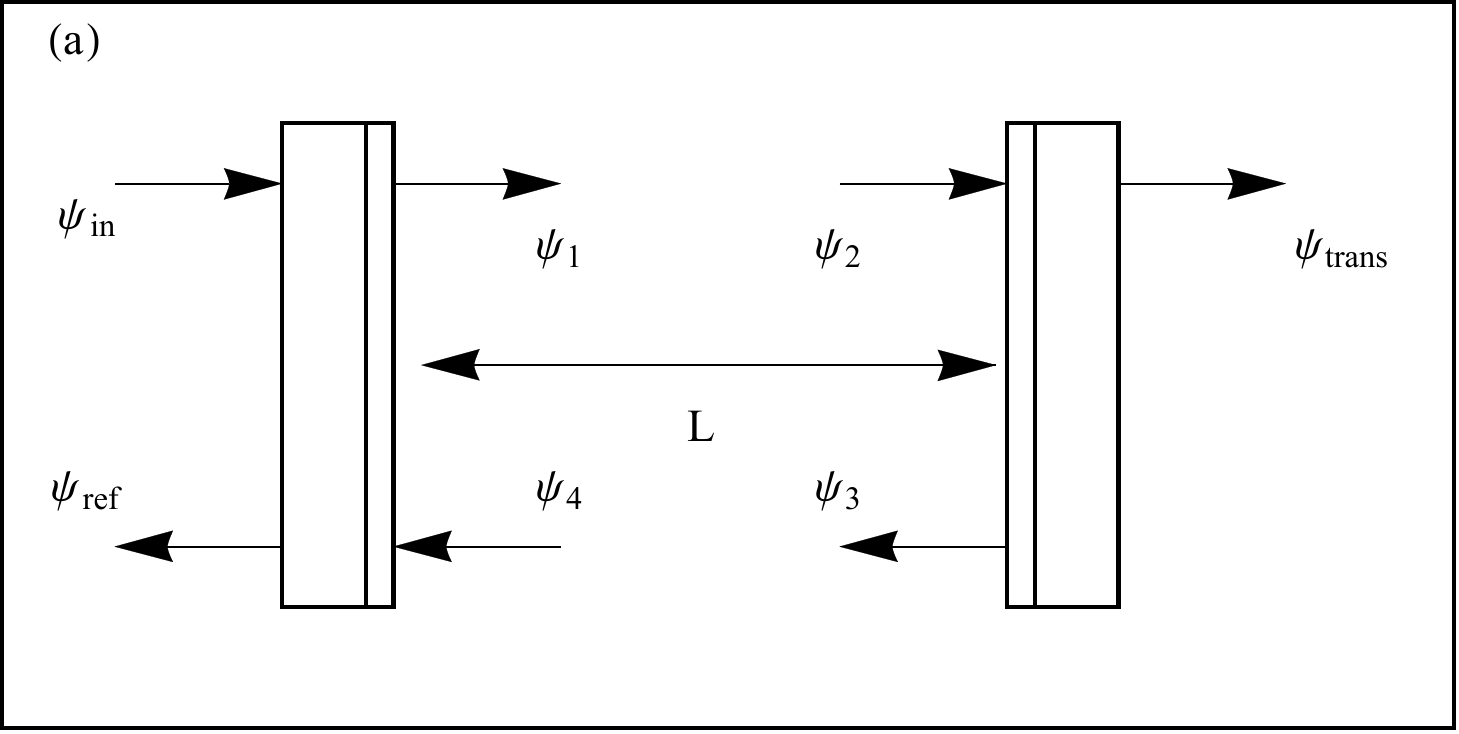}
$$ $$
\includegraphics[width=2.5in]{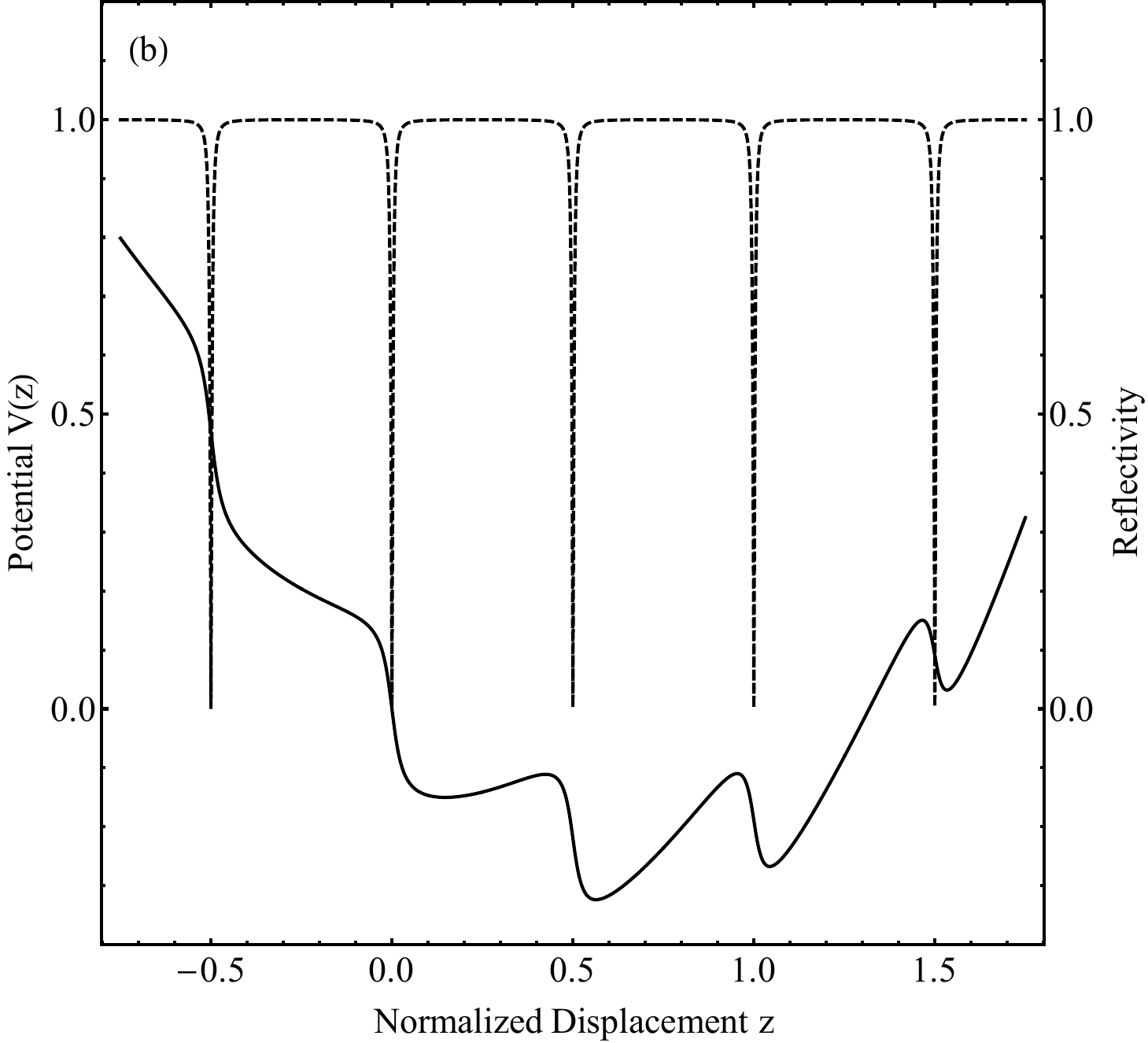}
\caption{ (a) the schematic of the mirrors $1$ (left) and $2$ (right). The incident electromagnetic field $\psi_{in}$ is partially 
reflected ($\psi_{ref}$) and partially transmitted ($\psi_{trans}$). 
The fields at the mirrors consequently read $\psi_1=t_1\psi_{in}-r_1\psi_4$, $\psi_{trans}=t_2\psi_1e^{-i\phi}$, $\psi_3=-\psi_2$, $\psi_{ref}=r_1\psi_{in}+t_1\psi_4$.
(b) the potential of the pendular FP interferometer (solid line) as a function of the displacement 
between the mirrors. The dashed line is the reflectivity, to be monitored to measure ETs. 
The minimum of the potential 
$V_-$ is marked by a black dot, while the minima of the reflectivity (indicating the escape) are 
marked by white dots at the potential $V_+$ and $V_+'$. 
The parameters of the system are: ${\cal F}=50$, $\Pi_{M}=5$, $t_1=t_2=0.97$, ${\cal L}=10^{-6}$.
${\cal L}$ is the loss factor.
}
\label{fig:potential}
\end{figure}

\section{Mathematical model}
\label{sect:model}

The pendulums for gravitational waves detection placed in a FP cavity, in absence of external deterministic signal is described by the following dimensionless equation:

\beq
\frac{d^2z}{dt^2} + \gamma \frac{dz}{dt} = - z + \Pi_{M}{\cal A}(2\pi z) + \xi(t),
\label{eq:zeta_norm}
\eeq
\beq
< \xi(t),\xi(t')> = 2D\delta(t-t').
\label{eq:correlator}
\eeq
\noindent

\noindent We denote with $z$ the displacement of terminal mass normalized respect to $\lambda$ , the wave length of incident laser radiation. 
The symbol $\xi(t)$ denotes  the random process incorporating the whole noise budget of the system applied on the end mass.
Furthermore  $r_i$ and $t_i$ are the reflection and transmission coefficients, 
$\Pi_{M} = R_{M}/\mu \lambda \omega_0^2$ is the normalized radiation pressure coefficient, $\mu$ and $\omega_0$ are the reduced mass of the pendulum and the mechanical frequency of the end mass, respectively. 
Time is normalized respect to the inverse of the natural frequency, $\omega_0$.
The normalized input power is $R_{M}= 2 P_{M}/c$ where $P_M$ is the laser input power.
The friction constant $\gamma=\tilde{\gamma}/\omega_0$ is given by the pendulum dissipation constant $\tilde{\gamma}$ divided by $\omega_0$ . 
The function ${\cal A}$ is the Airy function ${\cal A} =1/(1+{\cal F} \sin^2{\phi})$, where 
${\cal F}=\pi\sqrt{r_1 r_2}/(1-r_1r_2)$ is the Finesse of the cavity ($r_{i}$ ant $t_i$ are the reflection and transmission coefficents, respectively; $i=1,2$ refers to the mirros, see Fig. \ref{fig:potential}), and  $\phi=2\pi z$ is the phase of the circulating light.
The phases $\phi$ also determine the half-width $\delta \phi$ of the resonance, $\delta \phi = \pi/2 {\cal F}$. 
The maximum stored power corresponds to the peaks of the Airy function ($\phi=n\pi$). 
In Eq. (\ref{eq:correlator}) the noise intensity $D$ is assumed dominated by the external sources, and therefore does not obey the dissipation fluctuation theorem. 
This assumptions was also implicitly employed in Eq.(\ref{kramers}).
We underline that the dissipation parameter is very important. 
From the point of view of the detection, damping should be low enough to allow the cavity to reach high quality factors, and in fact actual systems achieve such small damping as $\gamma = 10^{-6}$\cite{Drever83}.

Equation (\ref{eq:zeta_norm}) describes the motion of a particle in the potential of Fig.\ref{fig:potential}b.
\beq
V(z)=
\frac{1}{2} z^2
- \frac{\Pi_M (
	\arctan (\sqrt{{\cal F}+1} \tan(2\pi  z)
	)+ \pi \lfloor 2 z+\frac{1}{2}\rfloor )}
{2 \pi 
   \sqrt{{\cal F}+1}}
\eeq
where $\lfloor \cdot \rfloor$ denote the floor truncation function.
The motion of this particle (without friction and external noise) can be also described by the hamiltonian
\beq
H(z,p)=
\frac{1}{2}p^2 + V(z)
\eeq
where $p$ is the particle momentum.

Placed in a minimum, the particle can escape subject to the random perturbation described by Eq.(\ref{eq:correlator}). 
In escaping, it crosses the points of sharp change of reflectivity, $V_+$ and $V_+'$. Thus the time elapsed to overcome the barrier, the escape time 
(or the first passage time \cite{Risken89}) given by Eq.(\ref{kramers}) can be detected without mechanical contacts. 

To evaluate the performances we have numerically retrieved the ETs distributions employing a quasi-symplectic modified velocity Verlet algorithm with velocity randomization for the integration of stochastic differential equations  \cite{Sivak12}. 
Moreover, we have  found consistent results  with another leapfrog algorithm \cite{Mannella04} that has proved to be very efficient even at extremely low dissipation \cite{Burrage07}. 
Finally, the algorithms have been tested against the known estimates for the washboard potential at very low dissipation ($\gamma << 1$) and noise ($D<<1$) \cite{Melnikov86}. We have found, for the whole dissipation range $  10^{-2}  > \gamma  > 10^{-6}$, that the theory lies within the $95\%$ confidence limit of the numerical simulations.

\begin{figure}
\centerline{\includegraphics[scale=0.4]{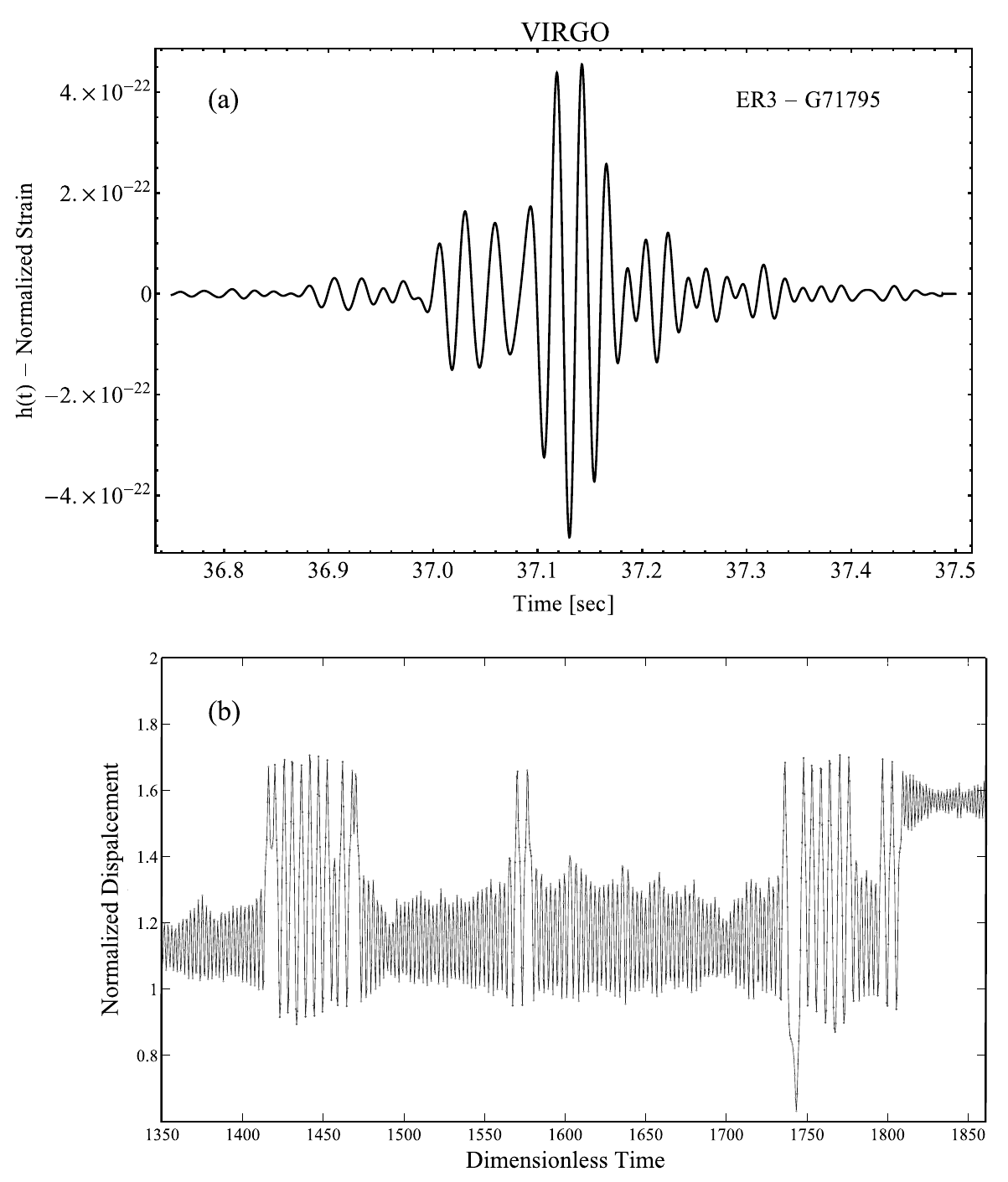}}
\caption{In (a) we display  pictorial time evolution for a glitch at Virgo, this is a signal extracted by the third Engineering Run (ER3)
labeled  G71795. 
In (b) we display a simulated transient in our system. 
The other parameters are $~{\cal F}=1000$, $D = 0.0005$, $\gamma = 10^{-2}$ and $\Pi_M=2$.
The similar behaviour suggests that in underdamped multi-stable systems, noise generates transitions
in nearly metastable states, that consequently produce large oscillations.
}
\label{fig:transient}
\end{figure}

Apart from the difficulty of the numerical simulations, low dissipation has the physical consequence of very long transients, on the scale $1/\gamma$. 
An example is shown in Fig. \ref{fig:transient}, for the real system (a) (see also \cite{Rampone13}) and simulations of Eqs.(\ref{eq:zeta_norm},\ref{eq:correlator}). 
The behavior is quite particular, in that the multiple metastable points of the potential (Fig. \ref{fig:potential})  can temporarily trap the pendulums. 
There are therefore several oscillations, with different amplitude, that can be activated by noise. 
It is particularly interesting that the oscillations at such diff rent scales are obtained by noise alone.

\section{Noise estimate}
\label{sect:thermal}
\begin{figure}
\centerline{\includegraphics[scale=0.35]{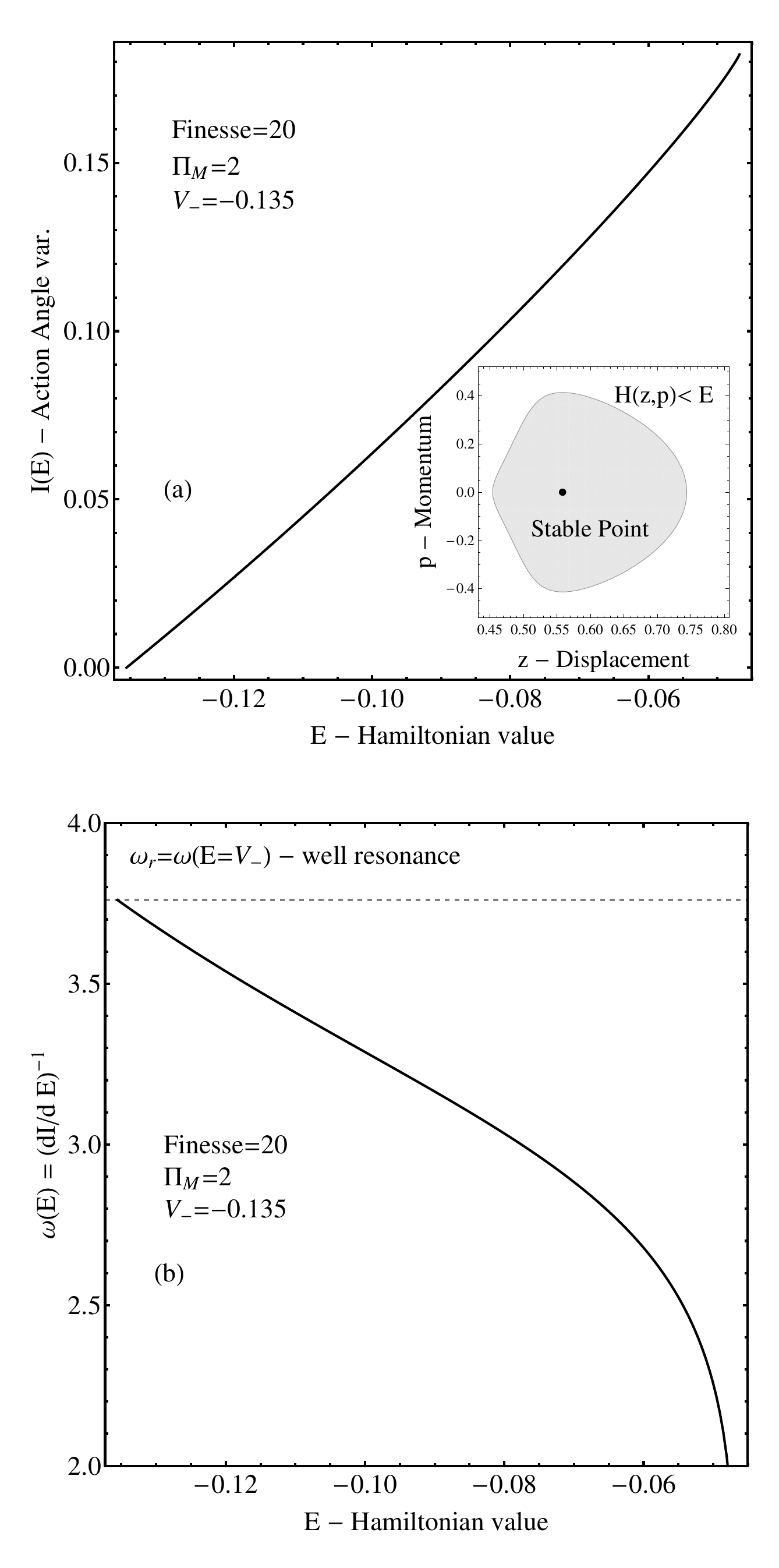}}
\caption{
Sketch of the evaluation of Eq.(\ref{eq:azan}) and of the conjugate $\omega(E)=(dI/dE)^{-1}$for ${\cal F} = 20$, $\Pi_M = 2$, $\Delta V = 0.042$.}
\label{Fig:analytical}
\end{figure}

The pendular interferometer, being characterized by a very low damping term $\gamma \simeq 10^{-6}$ \cite{Drever83}, 
should be handled in the extremely low dissipation limit. Some analytical approaches to handle activation processes as those entailed in 
Eq.(\ref{eq:zeta_norm})  have been proposed. 
These approaches extend the treatment of stochastic differential equations with vanishingly damping up to finite values \cite{Landauer83,Drozdov99}.
To apply the method described by Ref. \cite{Melnikov86}, we insert the total (electromagnetic and mechanical) potential 
 in the energy diffusion limit of the escape rate \cite{Mazo10}. 
In this approach the {\em action angle} variable  is used 
\beq
I(E)=\oint p dz
\label{eq:azan}
\eeq
to describe the diffusive equation,  the contour integral is defined by the isoenegy surface $H(z,p)=E$ (see inset in Fig. \ref{Fig:analytical}a).
A simple geometrical interpretation of eq. (\ref{eq:azan}) is that $I(E)$ is the area (in the phase space) of the closed surface
determined by the inequality $H(z,p)\leq E$ (as depicted in the inset of Fig.\ref{Fig:analytical}a).

Using the derivative of the action $I$ in the minimum of the potential $V_-$ and integrating by parts we obtain:
\begin{eqnarray}
r ^ {-1} &=& \frac{\gamma}{D^2}\int_{V_-}^{V_+} { I(E)e^{-E\gamma/D} dE \int_{E}^{V_+}
{\frac{e^{-E'\gamma/D}}{I(E')}  dE' }  } + \nonumber \\
&  +& \frac{\Delta V}{D}
\label{eq:integral}
\end{eqnarray}
Typical (numerical evaluation) of the $I(E)$ are shown in Fig.\ref{Fig:analytical}a, and the conjugate variable $\omega(E)=(dI/dE)^{-1}$ is reported
in Fig. \ref{Fig:analytical}b.
\noindent The advantage of formulation (\ref{eq:integral}) is that one can exploit the following parabolic approximation 
(by comparison with Fig.\ref{Fig:analytical}a it can  be found in acceptable agreement)
\beq
I(E) \simeq \frac{(V-V_-)2\pi}{\omega_r}+ \frac{1}{2}(V-V_-)^2 \left. \frac{d^2 I}{ dE^2}\right|_{E=V_{-}}
\eeq
to compute the average ET,  here $\omega_r$ is the well resonance and $\text{Ei}(\cdot)$ is the exponential integral function, 
see \cite{Prudnikov98}. 
The eq.(\ref{eq:integral}) can thus be analytically evaluated :
\[
r ^ {-1} =\frac{(1 +2 \rho)}{r_H}  -\frac{1}{\gamma} (1-2 \rho ) \log (1+\rho  \gamma \Delta V / D) +
\]
\beq
+~~\frac{1}{\gamma} e^{-1/\rho } (1+2 \rho) \left[\text{Ei}\left(\frac{1}{\rho }\right)-\text{Ei}\left(\gamma \Delta V / D+\frac{1}{\rho }\right)\right].
\label{eq:tau_theoretical}
\eeq
\noindent For 
\beq
\rho=  \frac{D \omega_r}{2\pi\gamma} \left. \frac{d^2 I}{ dE^2}\right|_{E=V_{-}}  = 0, 
\eeq
Eq.(\ref{eq:tau_theoretical}) reduces to the ET of the harmonic oscillator  (here $r_H$ is the rate for the harmonic oscillator and $\gamma_E$ is the Euler-Mascheroni constant):
\beq 
r^{-1}_H=\frac{1}{\gamma}\left[\text{Ei}\left(\frac{\gamma \Delta V}{ D}\right)-\log\left (\frac{\gamma \Delta V}{D}\right)-\gamma_E\right].
\eeq

More sophisticated  numerical evaluation of eq. (\ref{eq:integral})  based on the action-angle variable depicted in Fig. \ref{Fig:analytical}
introduce a relative error less than $1\%$ in any physical relevant situation.

\begin{figure}
\centerline{\includegraphics[scale=0.4]{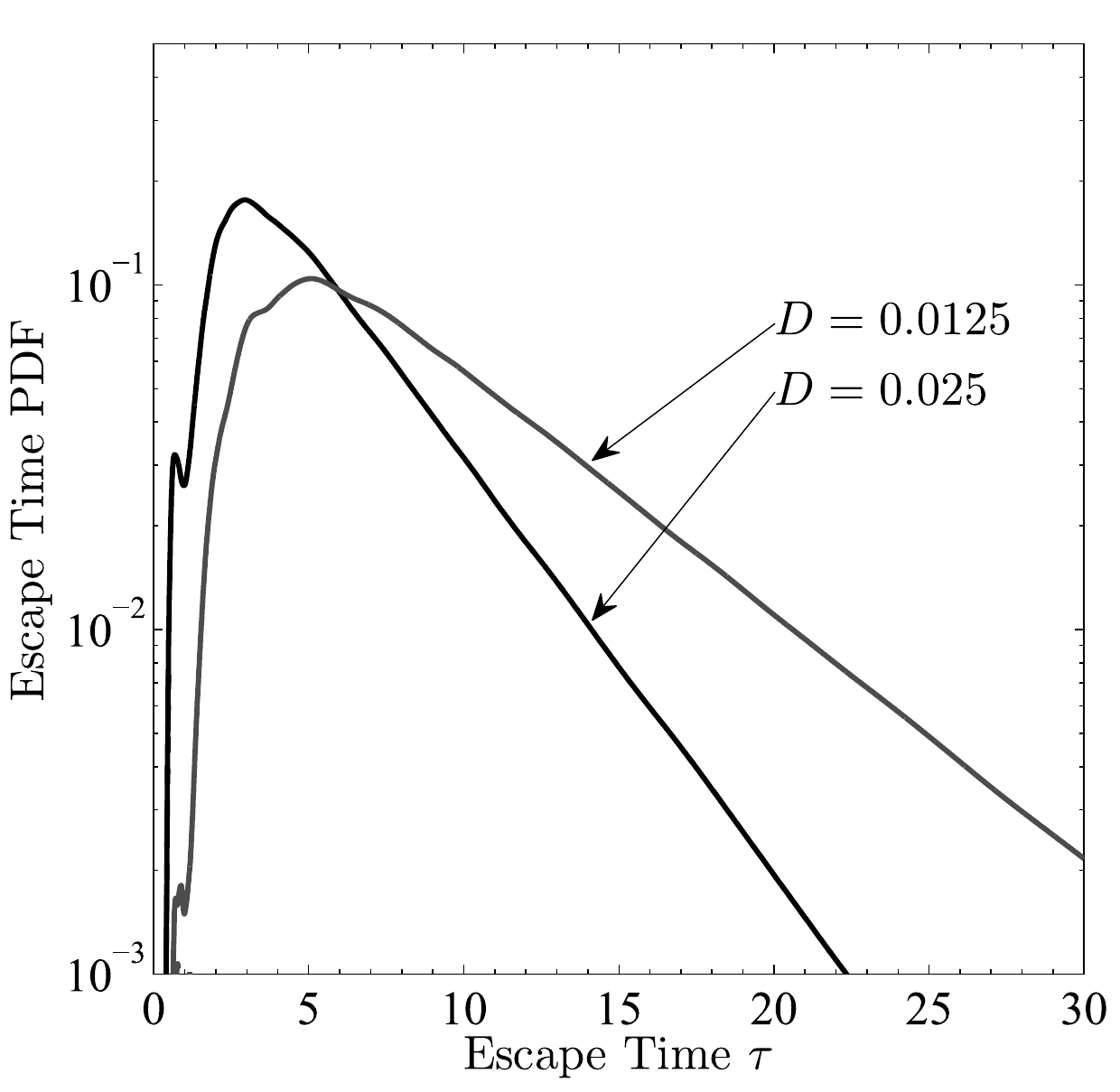}}
\caption{Probability distribution of the ET for two different values of the 
noise intensity ($D=0.025$ and $D=0.0125$).  The parameters of the system are: ${\cal  F} = 2$, $\Pi_{M} = 2.1$, $\Delta V = 0.07$, $\omega_r = 2.9$, $\gamma = 10^{-6}$.
}
\label{fig:PDF}
\end{figure}

The distributions of the ETs  shown in Fig.\ref{fig:PDF} is exponential (apart a cutoff) and it is evident that can be exploited to determine the intensity of the noise. 
 This amounts to inferring the noise level from the distributions of Fig. \ref{fig:PDF} with statistical estimation of the parameter $D$ in Eq.(\ref{eq:tau_theoretical}). 

\begin{figure}
\centerline{\includegraphics[scale=0.4]{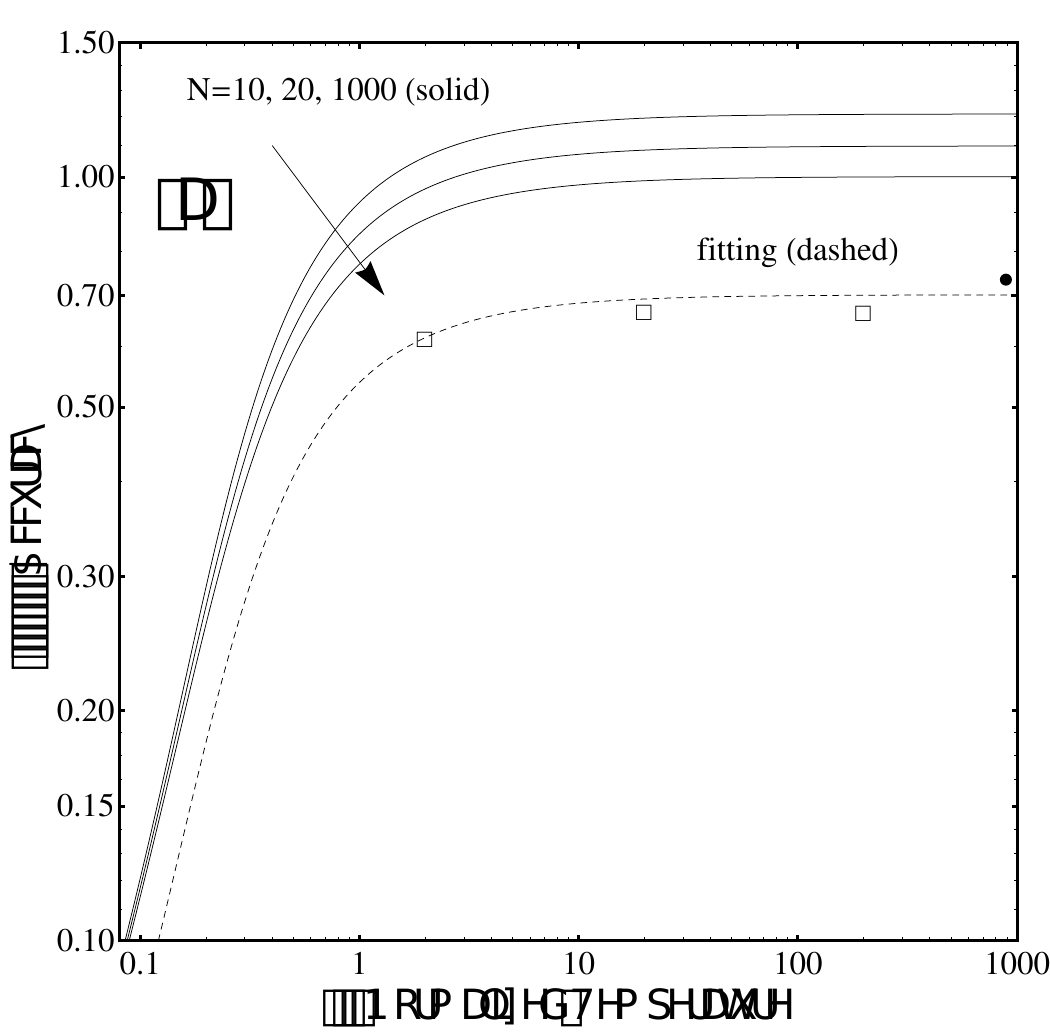}}
$$    $$
\centerline{\includegraphics[scale=0.43]{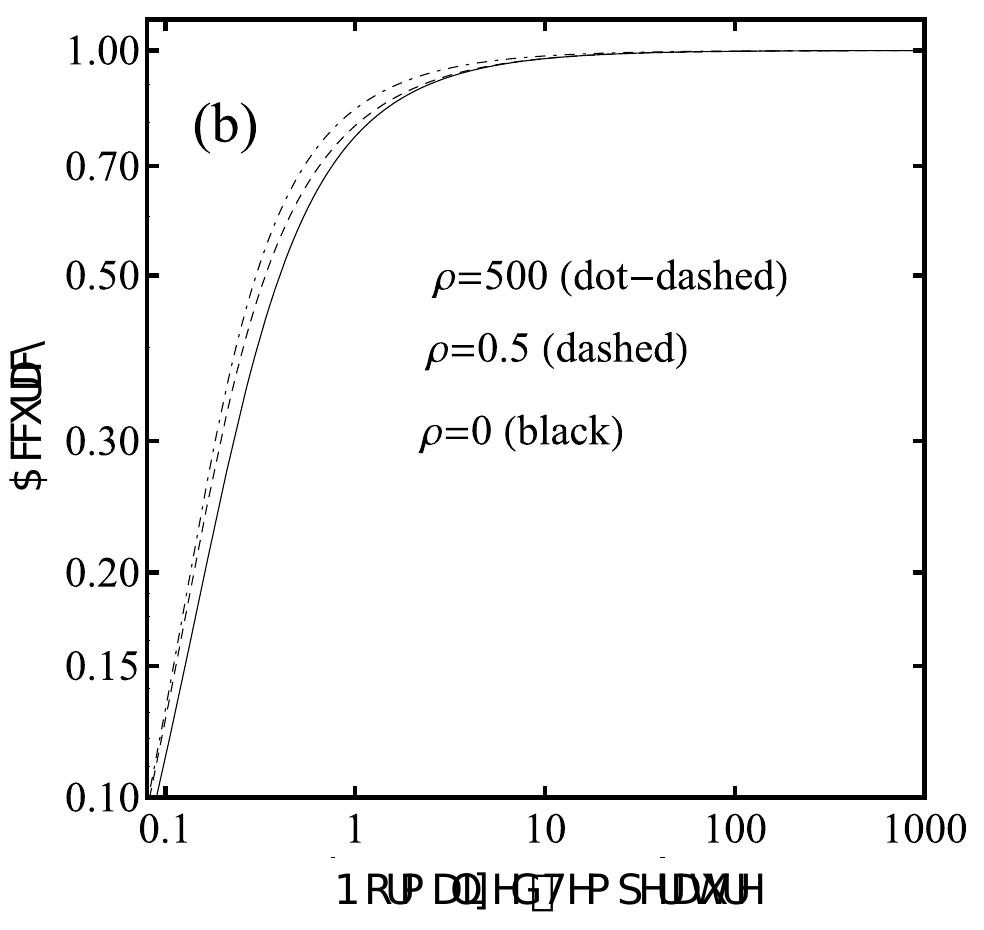}}
\caption{
(a) Relative accuracy $\sqrt{N}\sigma_N/D$ of the detected noise level as a function of the normalized external noise intensity $D/(\gamma\Delta V)$ for different values of the sample numerosity. 
Symbols refer to simulations of the Langevin Eq.(\ref{eq:zeta_norm}) for $N=1000$, ${\cal F} = 20$, $\Pi_M = 2$, $\Delta V = 0.042$, $D=0.00042$ (filled points) and $D=0.00084$ (empty circles). The dashed line denotes the fitting. 
(b) The asymptotic  behaviors as a function of the parameter $\rho$ that represents the deviations of the model Eq.(\ref{eq:zeta_norm}) from the harmonic oscillator ($\rho=0$).}
\label{fig:noisedetection}
\end{figure}

The efficacy of the method for the estimate of the noise level has been evaluated computing the  variance $\sigma^2_N$ of the estimated temperature as a function of the sample size $N$ and 
of the noise intensity $D$, under the hypothesis that ETs are described by an exponential distribution.
One can derive the large $N$ behavior of the Maximum Likelihood estimator for an exponential distribution; in fact the Fisher information $J$ of the distribution given by Eq.(\ref{eq:tau_theoretical}) reads: $\sigma_N^2 \simeq J^{-1}/N$ \cite{Lehmann99}. 
Thus, the theoretical estimate shows that the relevant parameter is the noise $D$ normalized to the product of damping and the energy barrier, $\gamma \Delta U$.
The dependance is displayed in Fig. \ref{fig:noisedetection}.

The behavior is therefore shown in Fig. \ref{fig:noisedetection} as a function of the noise temperature. The variance of the estimate relative error increases with noise level up to a saturation point (around $D\simeq \gamma \Delta V$) that depends, as expected, upon $N^{-1/2}$ and exhibits a weak dependence upon $\rho$ (see \ref{fig:noisedetection}).
The estimate of the noise level improves by increasing $\gamma\Delta V$ i.e. 
the dissipation and/or the potential barrier. From the physical consideration that a minimum time occurs before escape \cite{Berglund05} (see also Fig. \ref{fig:PDF}), it is evident that the ET density departs from the exponential model for short escapes. 
Such deviations from the exponential distribution lead to an overestimate of the relative error (as evident from the fitted curve in Fig. \ref{fig:noisedetection}) numerically evaluated to be about $30\%$ smaller.

Apart the finite size error, the method could also be biased. This has ben checked by numerical simulations, and we have found that the distortion of the temperature estimate is below $5\%$.

Finally, we will hint to the effect of a periodic drive, to mimic the perturbation of a gravitational wave. 
If such a sinusoidal term is added in Eq.(\ref{eq:zeta_norm}), the distribution of the ETs is modified \cite{Berglund05}and  the ETs can be used for the signal detection \cite{Addesso13}.

\section{Conclusion}
We propose to employ the abrupt reflectivity changes to estimate the noise level of pendular FP.
 We have shown that an effective estimate of the noise can be reached. Realistic models include very low dissipation $\gamma \simeq 10^{-6}$. 
 Such low damping, combined with  multistable potential determines a peculiar behavior of the transient , that exhibit oscillations at different amplitudes. 
 This could resemble glitches, as shown in Fig. \ref{fig:transient}. 
We have also estimated, in the very low damping limit, the average escape rate. 
Assuming an exponential distribution, the approximation allows for an analytical evaluation of the reliability of the noise estimate. 
In particular, as also demonstrated by the simulations, we can ascertain that the estimate improves as the square root of the number of detected escapes, as in optimal estimates. The analysis goes further, and gives also the behavior as a function of the system parameters. 
While noise estimate is a preliminary measurement, the natural question to follow is about signal detection. It has been proved that resonant activation is connected to some sub-optimal strategies, while for optimal strategies it disappears \cite{Addesso13}. 
In perspective we propose that it could be interesting to estimate the noise level when the disturbance is not Gaussian, as expected at low frequency as a consequence of ground vibrations and fluctuations of the electronics.
By way of conclusion, the preliminary analysis indicates that it is promising to exploit the reflectivity changes in that the measurements are mechanically decoupled, but the loss of information is relatively mild.
\[~
\]
\section*{Acknowledgment}
The authors would like to thank Prof. I. M. Pinto for illuminating discussions.
\vspace{13cm}
$$~$$

\end{document}